\documentclass[12pt]{article}

\usepackage{epsfig}
\usepackage{graphicx}
\usepackage{color}
\usepackage{soul}
\usepackage{times}
\usepackage[superscript]{cite}

\begin{document}
\title{Dual behavior of excess electrons in rutile TiO$_2$}

\author
{A. Janotti,$^{1\ast}$ C. Franchini,$^{2}$ J. B. Varley,$^{1}$ \\
G. Kresse,$^{2}$ C. G. Van de Walle$^{1\ast}$\\
\\
\normalsize{$^{1}$Materials Department, University of California, }\\
\normalsize{Santa Barbara, CA 93106-5050, USA}\\
\normalsize{$^{2}$Faculty of Physics, University of Vienna and }\\
\normalsize{Center for Computational Materials Science, A-1090 Wien, Austria}\\
\\
}

\baselineskip24pt


\maketitle

The behavior of electrons in the conduction band of TiO$_2$ and other transition-metal oxides is key to the many applications of these materials.
Experiments seem to produce conflicting results:
optical and spin-resonance techniques reveal strongly localized small polarons, while electrical measurements show high mobilities that can only be explained by delocalized free electrons.
By means of hybrid functional calculations we resolve this apparent contradiction and
show that small polarons can actually coexist with delocalized electrons in the conduction band of TiO$_2$, the former being
energetically only slightly more favorable.
We also find that small polarons can form complexes with oxygen vacancies and ionized shallow-donor impurities, explaining the rich spectrum of Ti$^{3+}$ species observed in electron spin resonance experiments.


\newpage

The performance of transition-metal oxides such as TiO$_2$ in (photo)catalysis, photosensitized solar cells, and memristors
is tightly linked to the properties of conduction-band electrons.
The envisioned applications of these oxides as semiconductors for next-generation electronics lends particular urgency to the development of a deeper understanding of conduction mechanisms \cite{Jang11}.
Seemingly conflicting results have been reported for the behavior of excess electrons in TiO$_2$:
While high mobilities have been observed \cite{EY96}, characteristic of delocalized electrons, excess electrons in TiO$_2$ have long been described as localized small polarons \cite{Mott01,Ston07,Eag64}.
The current understanding of the optical and electronic-transport properties in TiO$_2$ is thus based on ostensibly contradictory physical scenarios:

\noindent  (i) $n$-type conductivity can be readily achieved by incorporation of donor impurities such as Nb, F, or H or by annealing in vacuum at high temperatures; electron mobilities as high as  $10^3$ cm$^2$V$^{-1}$s$^{-1}$ at 20 K have been reported \cite{EY96,Mott01}.

\begin{figure}
\center
\includegraphics[width=0.6\columnwidth]{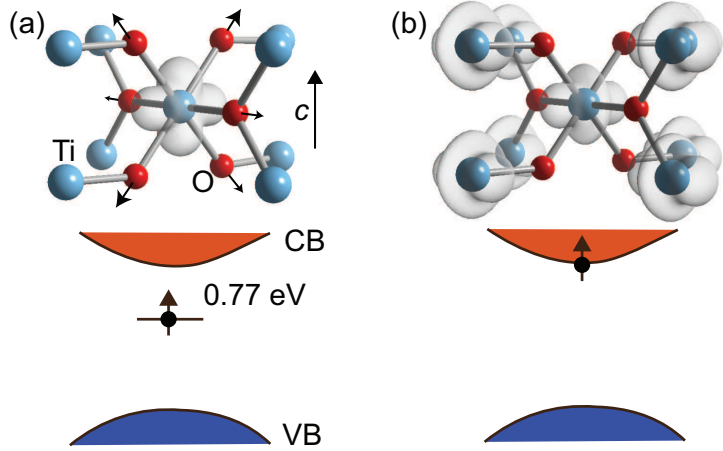}
\caption{Charge distribution of a small polaron and a delocalized conduction-band electron in TiO$_2$. (a) Spin density of a self-trapped electron (small polaron) corresponding to a single-particle level at 0.77 eV below the conduction band in rutile TiO$_2$. (b) Charge distribution of a delocalized electron in the conduction band in TiO$_2$. In both cases, the isosurfaces correspond to 10\% of the maximum charge density. The corresponding band structures are shown underneath, where CB and VB refer to valence band and conduction band, respectively.
\label{fig:fig1}}
\end{figure}

\noindent (ii) At the same time, the electrons provided by shallow-donor impurities or native point defects (O vacancies or Ti interstitials) have been assumed to localize on individual Ti lattice atoms. This small polaron, consisting of a localized electron and the accompanying local lattice distortion, gives rise to an optically detected deep level at 0.8 eV below the conduction band (Fig.~\ref{fig:fig1}).  This level has been characterized by infrared and core-level x-ray spectroscopy and by electron spin resonance (ESR), and has commonly been assigned to Ti$^{3+}$ centers \cite{Mott01,Ston07,Eag64,BG68,Cron59,Hall09}.

\noindent (iii) The role of native point defects in the electrical properties of TiO$_2$ is a matter of debate. Some experiments indicate that oxygen vacancies can trap one or two electrons, resulting in $S$=1/2 and $S$=1 spin configurations with magnetic moments localized on neighboring Ti$^{3+}$ \cite{Ston07,Hall09,Hall10}, while others claim that oxygen vacancies are shallow donors, responsible for the $n$-type conductivity observed in vacuum-annealed samples \cite{EY96,Nowo06,AONO93,K64}.

\noindent (iv) To complicate the picture even further, results of {\em ab initio} calculations by different groups strongly disagree on whether oxygen vacancies are deep donors, and hence electrically inactive, or shallow donors responsible for $n$-type conductivity \cite{JMS03,SNP06,Iddir07,He07,LZ07,Pacch08,GWW09,AJ10}.

These inconsistencies motivated us to investigate the problem by
performing hybrid functional calculations.
Excess electrons in TiO$_2$ occupy narrow bands derived from Ti $d$ orbitals and strongly interact with phonons in a highly polarizable lattice.
This requires the use of a computational approach which can accurately describe localization effects, in addition to reliably predicting
structural relaxations, energetics and band structure.
Hybrid functionals \cite{hse} within density functional theory have been shown to have this capability \cite{Franchini09,Varley12}.
Our results provide a consistent and unified physical picture.
We find that self-trapped electrons (small polarons) are only slightly lower in energy than delocalized electrons in the conduction
band ($\Delta$E=0.15 eV). The calculated optical transition level (0.61 eV) is in good agreement
with the observed absorption peak at 0.8 eV.   We also calculate a migration barrier of 0.03 eV for the small polaron, which is too high
to explain the observed electron mobility of $10^3$ cm$^2$V$^{-1}$s$^{-1}$ at 20 K in TiO$_2$ single crystals \cite{EY96};
the observed conductivity must therefore be attributed to delocalized electrons in the conduction band.

Our study also resolves the long-standing controversy regarding the electronic structure of oxygen vacancies in TiO$_2$:
the vacancy is intrinsically a shallow donor, but small polarons can bind to it, similar to the binding of conduction-band electrons
to shallow donors in hydrogenic effective mass states.
Likewise, small polarons can form complexes with impurities such as Nb, F, and H, which also act as shallow donors.
The formation of Ti$^{3+}$ near a defect or impurity was recently addressed in density functional calculations based on the B3LYP functional \cite{Valentin09} and LDA+$U$ \cite{Morgan10,Mattioli10}; however, those studies did not address small polarons as an intrinsic property of TiO$_2$.
Small polarons in defect-free TiO$_2$ have been described with LDA+$U$ calculations \cite{GWW09,Deskins}, but those studies did not resolve the issue of the stability of polarons versus delocalized electrons since the latter cannot be properly described within LDA+$U$.

The calculations are based on generalized Kohn-Sham theory within the projector-augmented wave method as implemented in
the VASP code \cite{dft,kressepaw,vasp1,vasp2}.
We use the hybrid functional of Heyd, Scuseria, and Ernzerhof
(HSE) \cite{hse}, in which the exchange and correlation potential is divided into long- and short-range parts.
Non-local Hartree-Fock exchange is mixed with the semi-local exchange of Perdew, Burke, and Ernzerhof \cite{pbe} (PBE) in the short-range part, while the correlation and the long-range part of the exchange potential are described by PBE.
The mixing parameter and the inverse screening length were set to the standard values of 0.25 and 0.1 \AA$^{-1}$, respectively. Spin polarization was included. We use a cutoff energy of 300 eV for the plane-wave basis set.  Convergence checks were performed using 72-, 96-, and 216-atom supercells with a 2$\times$2$\times$2 set of k-points, or $\Gamma$ only in the 216-atom cell.  The results reported here are for the 216-atom supercell.

Small polarons are modeled by studying the self-consistent electronic and structural response of the system to an extra electron added to the conduction band of an otherwise perfect TiO$_2$ crystal.
We find that the excess electron leads to two distinct and locally stable solutions: (i) the electron occupies an extended state at the bottom of the conduction band, and (ii)
the electron occupies a localized state centered on an individual Ti site, forming a Ti$^{3+}$ center.  The calculated spin density for the localized, self-trapped electron is shown in Fig.~\ref{fig:fig1}(a).
The electron localization is accompanied by a distinct local lattice distortion around the Ti$^{3+}$ ion: the two
in-plane (perpendicular to the $c$ axis) nearest-neighbor O atoms relax outward by
1\% of the equilibrium Ti-O bond length, and the four out-of-plane O atoms by 4\%.
This quasiparticle consisting of an electron and the surrounding polarization field is a small polaron.
The electron is localized primarily on a particular Ti atom, with the corresponding $t_{2g}$-like {\em single-particle} state
positioned at 0.77 eV below the conduction band.
For comparison, in Fig.~\ref{fig:fig1}(b) we show the spin density for a delocalized electron in the TiO$_2$ conduction band. In this case the
electron density is equally distributed over all Ti atoms of the undistorted TiO$_2$ lattice.

\begin{figure}
\center
\includegraphics[width=0.6\columnwidth]{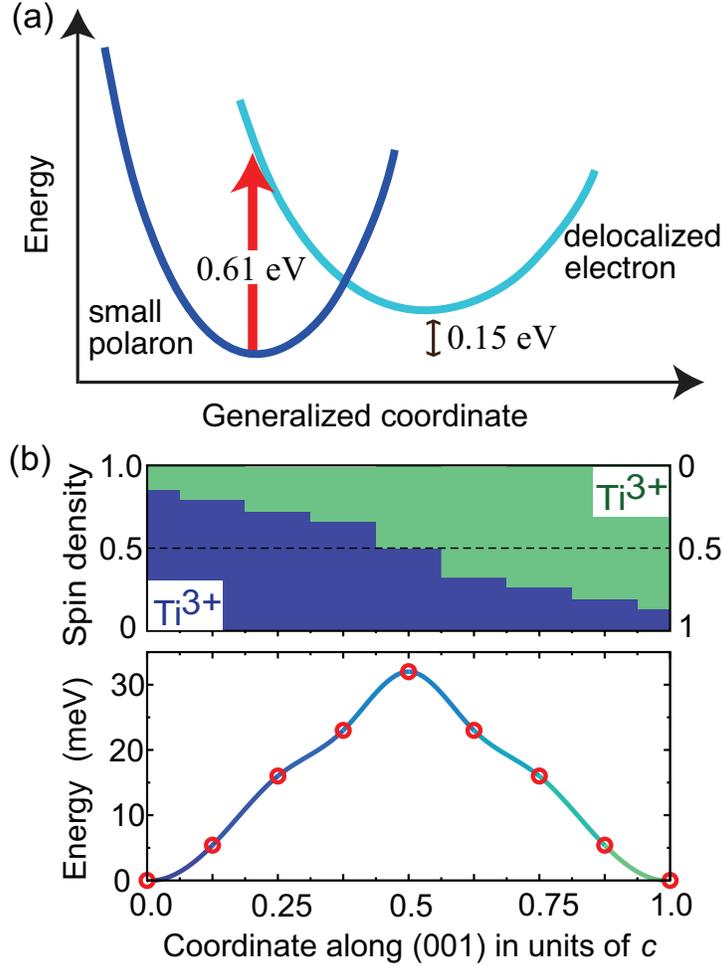}
\caption{Excitation and migration mechanism for a small polaron in TiO$_2$.
(a) Calculated configuration coordinate diagram depicting the energy as a function of lattice distortion for a polaron (left) and a delocalized electron (right), with an energy difference between ground-state configurations of 0.15 eV.  Absorption of a photon with energy 0.61 eV causes an optical transition from a polaron to a delocalized electron.
The value of 0.61 eV was obtained by adding the 0.15 eV energy difference between the perfect lattice with an electron at the bottom of the conduction band and the polaron, to the energy required to distort the perfect lattice to the polaronic configuration, 0.46 eV.
(b) Potential energy for a polaron migrating along the $c$ axis, along with the spin density projected on the two neighboring Ti atoms (using a radius of 1.32 \AA) as the polaron is transferred from one (left) to another (right).
\label{fig:fig2}}
\end{figure}

The stabilization of the small-polaron configuration can be understood as resulting from the energy balance between the electronic energy
gained by placing the electron in a localized $d$-state 0.77 eV below the conduction band (Fig.~\ref{fig:fig1}(a)),
and the strain energy required to distort the lattice and accommodate the larger Ti$^{3+}$ ion.
We can calculate this strain energy as the total-energy difference between the distorted lattice and the perfect lattice in the absence of the excess electron, resulting in a value of 0.46 eV.  These results allow us to construct the configuration coordinate diagram shown in Fig.~\ref{fig:fig2}(a).
By adding the total-energy difference between the polaronic (Fig.~\ref{fig:fig1}(a)) and the delocalized-electron (Fig. \ref{fig:fig1}(b)) configuration (0.15 eV) to the strain energy (0.46 eV), we obtain a vertical excitation energy of 0.61 eV, which is in good agreement with the reported infrared absorption peak at 0.8 eV, commonly assigned to a  transition involving an electron from a Ti$^{3+}$ state to the conduction band \cite{BG68,Eag64}.
This optical transition energy is also consistent with the energy of the single-particle state (0.77 eV below the conduction band) obtained from the band-structure calculation.  It is also significantly larger than the 0.15 eV energy difference between the electron in the small-polaron and delocalized-electron configurations, with important repercussions for optical {\it versus} transport measurements of excess electrons in TiO$_2$.

We now address whether the observed electron conductivity in $n$-type TiO$_2$ can be reconciled with the existence of small polarons.
We calculated the migration path of a polaron by interpolating the atomic positions of two
adjacent and equivalent polaronic configurations, corresponding to an adiabatic process \cite{Deskins}.
For each intermediate configuration, the electronic structure was solved self-consistently.
The variation of total energy as a function of the fraction of initial and final atomic positions is plotted in Fig. \ref{fig:fig2}(b).  We also show the spin density projected on the two adjacent Ti atoms as the localized electron is transferred from one Ti atom to another.
For migration along the $c$ axis, in which the neighboring Ti sites are separated by 2.95 \AA,
we find an energy barrier of 0.03 eV (Fig. \ref{fig:fig2}(b)).
The energy barrier for migration along the [111] direction, accounting for transport perpendicular to the $c$ direction, is 0.09 eV, i.e.,  almost three times higher than along the $c$ direction.  Using the Einstein model for diffusion \cite{Einstein}, with a prefactor based on the optical-phonon frequency in TiO$_2$ ($\sim10^{13}$ s$^{-1}$), the resulting polaron mobility at 20 K is $\sim$10$^{-7}$ cm$^2$V$^{-1}$s$^{-1}$.
This value is many orders of magnitude smaller than the carrier mobilities of 10$^3$ cm$^2$V$^{-1}$s$^{-1}$ extracted from Hall measurements on TiO$_2$ single crystals \cite{EY96}.
An LDA+$U$ study \cite{Deskins} produced a higher value for the polaron migration barrier than our present result, which would imply even lower mobilities.   Moreover, small-polaron hopping is a thermally activated process and would lead to an increase in mobility as the temperature increases, contrary to the experimental observations \cite{EY96}.  The experimentally observed mobilities are therefore clearly not compatible with a small polaron hopping process, and can only be explained by invoking transport via delocalized electrons in the conduction band.

But how to reconcile this with our finding that the polaronic state is more stable (by 0.15 eV, Fig.~\ref{fig:fig2}(a))? Thermal excitation is insufficient to excite a significant number of electrons to the conduction band at a temperature of 20 K. However, we find that there is an energy barrier (Fig.~\ref{fig:fig2}(a)) that confines an electron to its delocalized, metastable state and keeps it from accessing the small-polaron regime.  This energy barrier was determined by self-consistent total energy calculations of various lattice configurations that were interpolated between the polaronic and the free-electron configurations. This procedure likely results in an underestimation of the actual barrier for a small polaron to convert into a delocalized electron, as it requires a concerted motion of atoms. This barrier will prevent the delocalized electron from collapsing into the polaronic state. In Hall or conductivity measurements, electrons are injected from metal contacts into extended conduction-band states and travel as delocalized carriers giving rise to high electron mobility.   These delocalized electrons of course experience phonon scattering as in conventional semiconductors \cite{Singh}, explaining the decrease in mobility as the temperature increases.

We now turn to the issue of interactions between small polarons and donor centers in TiO$_2$.
Considering that these small polarons are negatively charged centers, we expect them to be attracted to
ionized donor defects or impurities.
Indeed, we find that donors can form complexes with small polarons, in which the electrons are localized on neighboring Ti sites (Ti$^{3+}$).
These complexes are characterized by binding energies in the range of 0.10 to 0.20 eV, depending on the impurity or defect.
Recently we reported that the oxygen vacancy $V_{\rm O}$ in rutile TiO$_2$ is stable exclusively in the +2 charge state, with
the neutral and +1 charge states higher in energy for Fermi-level positions within the band gap \cite{AJ10}.

\begin{figure}
\center
\includegraphics[width=0.6\columnwidth]{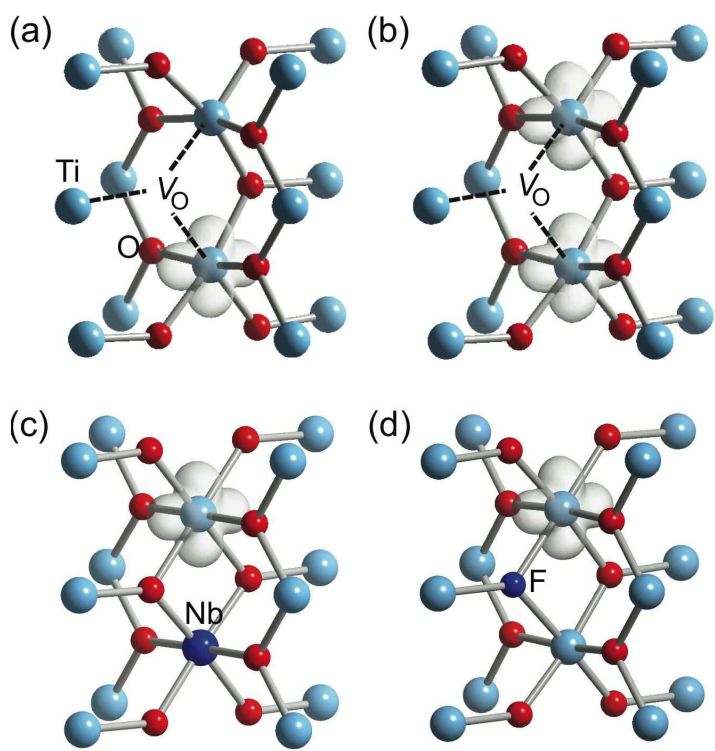}
\caption{Interaction between small polarons and shallow-donor centers in TiO$_2$.
(a) Spin density of a complex between an ionized oxygen vacancy ($V_{\rm O}^{+2}$) and a nearest-neighbor small polaron.
(b) Spin density of two polarons bonded to  $V_{\rm O}^{+2}$.
(c) and (d) Spin densities of a polaron bonded to Nb$_{\rm Ti}^+$ and F$_{\rm O}^+$.  All isosurfaces correspond
to 10\% of the maximum densities.
\label{fig:fig3}}
\end{figure}

As shown in Figs. \ref{fig:fig3}(a) and \ref{fig:fig3}(b), we find that $V_{\rm O}^{+2}$ can trap one or two polarons, with $S$=1/2 and $S$=1, respectively.
We note that these configurations are very different from those of $V_{\rm O}^{+}$ and $V_{\rm O}^{0}$ reported in Ref.~\cite{AJ10}, defined as configurations in which the electrons occupy a defect state whose density is centered on the O vacant site.  The Ti-Ti distances in the complex with two polarons bound to $V_{\rm O}^{+2}$ are the same as in the isolated $V_{\rm O}^{+2}$, and significantly larger than the Ti-Ti distances in $V_{\rm O}^{0}$ from Ref.~\cite{AJ10}.
In addition, the spin state of $V_{\rm O}^{0}$ (in which two electrons are localized on the point defect itself) is $S$=0, while $V_{\rm O}^{+2}$ plus two polarons has $S$=1.  The $S$=1 signal is what is observed in ESR measurements \cite{Hall09}.
We conclude that while both $V_{\rm O}^{0}$ (as defined in Ref.~\cite{AJ10}) and the complex consisting of $V_{\rm O}^{+2}$ plus two polarons are locally stable and overall charge-neutral, they have very distinct physical characteristics, and it is the ($V_{\rm O}^{+2}$ + 2 polarons) complex that is more energetically favorable, resulting in shallow-donor behavior of the vacancy.
The binding energy is 0.14 eV for the first polaron and 0.10 eV for the second.
These binding energies are small enough to allow some fraction of the complexes to be thermally dissociated at room temperatures,
resulting in ``free'' polarons or delocalized electrons in the conduction band.

Intricate structural arrangements for the ``neutral'' vacancy  have been reported based on {\em ab initio} calculations \cite{Valentin09,Morgan10,Mattioli10}.
Those structures agree with our results for $V_{\rm O}^{+2}$ plus two polarons shown in Fig. \ref{fig:fig3}(b) -- not $V_{\rm O}^{0}$, which is higher in energy \cite{AJ10}.
Our results are in line with recent hybrid functional calculations (Ref. \cite{Deak12}, published after the submission of the present work).
Previous studies \cite{Valentin09,Morgan10,Mattioli10} did not distinguish the charge-neutral complex consisting of $V_{\rm O}^{+2}$
plus two polarons from the vacancy in the neutral charge, thus failing to recognize that the formation of small polarons is an intrinsic property of the host material,
and their existence does not depend on the specific nature of the shallow donor centers.
Indeed, we find that polarons also form complexes with ionized donor impurities such as
Nb substituting on a Ti site (Nb$_{\rm Ti}$) or F substituting on an O site (F$_{\rm O}$),
as also observed in GGA+$U$ \cite{Morgan09} and hybrid functional calculations \cite{Deak11,Yamamoto12}.
The spin densities of the corresponding neutral complexes are shown in Figs. \ref{fig:fig3}(c) and \ref{fig:fig3}(d).
The calculated binding energies are 0.09 eV for Nb$_{\rm Ti}^+$ and 0.20 eV for F$_{\rm O}^+$.
At sufficiently low temperatures small polarons will thus be trapped at donor centers.
These complexes give rise to optical absorption in the range 0.8-1.1 eV,
i.e., the sum of the transition energy for the polaron and the binding energy to the donor center.
This result explains the experimental findings that at low temperatures the infrared absorption peak due to polarons shifts to higher energies, depending on the donor impurity  \cite{BG68,Cron59}.

The insights about the dual behavior of electrons produced by our hybrid functional calculations should stimulate new experimental studies as well as a reinterpretation of results already in the literature.  Polarons bound to shallow donors (be it oxygen vacancies or dopant impurities) should be viewed as {\it complexes} rather than configuration of the donor in a neutral charge state, and binding energies can be obtained from temperature-dependent measurements.  The high mobilities observed at low temperatures merit deeper investigation to identify the conditions under which electrons can be prevented from transforming into small polarons, as well as the time and length scales on which this process occurs.  Such insights could prove essential to the utilization of transition-metal oxides in novel electronic devices.

\section*{Acknowledgements}

We gratefully acknowledge R. A. Street and the late A. M. Stoneham for fruitful discussions.
This work was supported by the NSF MRSEC Program  (DMR-1121053),
by the U. S. Army Research Office (W911-NF-11-1-0232),
and by the Austrian FWF.  Computing resources were provided by
the Center for Scientific Computing from the CNSI/MRL under NSF MRSEC (DMR-1121053)  and NSF CNS-0960316,
XSEDE (DMR-070072N) which is supported by NSF grant number OCI-1053575,
and the Vienna Scientific Cluster under the EU-FP7 grant ATHENA.

\end{document}